\documentstyle[preprint,aps]{revtex}
\begin{document}
\draft
\title{Temperature dependence of the nematic anchoring energy:\\
mean field analysis}
\author{G. Barbero$^1$ and A. K. Zvezdin$^{1,2}$}
\address{$^1$ Dipartimento di Fisica del Politecnico di Torino\\
and Istituto di Fisica della Materia, Unita' di Ricerca Torino Politecnico\\
Corso Duca degli Abruzzi 24, I-10129 Torino, Italia\\
$^2$ Institute of General Physics of the Russian Academy of Sciences\\
Vavilov Street 38, 117942 Moscow, Russia}
\date{\today}
\maketitle
\begin{abstract}
In the mean field approximation, we evaluate the temperature dependence
of the anchoring energy strength of a
nematic liquid crystal in contact with a solid substrate due to
thermal fluctuations.
Our study is limited to
the weak anchoring case, where the microscopic surface energy is 
small with respect to the mean field energy due to the nematic phase. We assume
furthermore that the physical properties of the substrate can be
considered temperature independent in the
range of the nematic phase. 
According to the thermodynamical
perturbative approach, the macroscopic surface energy is deduced
by averaging the microscopic one, with a density matrix containing only
the nematic mean field. We show that the thermal renormalization of
the anchoring energy coefficients is proportional to
the generalized nematic order parameters. 
Our analysis shows also that
the thermal renormalization of the anchoring energy coefficients
predicted by means of Landau-like theories is a first and rather
rough approximation in the whole nematic temperature range. 
\end{abstract}
\pacs{61.30.-v,61.30.Cz,68.10.Cr}
\narrowtext  
One of the least-understood areas of physics and chemistry of liquid crystals
concern the anchoring phenomenon and the temperature surface transitions
at the interface liquid crystal-solid or soft substrate. These phenomena
are important also from a practical point of view, since they play
fundamental role in the realization of displays. In this Letter we analyse
these phenomena in nematic liquid crystal media.

Nematic liquid crystals are anisotropic fluids made by anisometric
molecules having quadrupolar symmetry.  Their intermolecular interaction,
$V_N$, is such to orient the
molecular axes, ${\bf u}$, along a common direction ${\bf n}$, called
director~\cite{deGennes}. 
It coincides with the optical axis of the medium. When a nematic liquid
crystal is in contact with a substrate, the orientation of ${\bf n}$ at
the surface results from a balance of the anisotropic 
interactions  with the bulk and with the substrate. In
the absence of bulk distortions, the surface orientation of the nematic
director coincides with the ``easy axis", ${\bf n}_0$~\cite{Sonin}. It is such to minimize the anisotropic
part of the surface energy characterizing the interface between the
nematic and the substrate. Long ago Bouchiat and
Langevin-Cruchon~\cite{Bouchiat} found a strong temperature dependence
of the easy axis. The measurements of Ref.~\cite{Bouchiat} have
been repeated by other groups with similar
results~\cite{FC1,FC2,R3,L,Patel,M}.
Several models have been proposed to interpret
this phenomenon. According to Parson~\cite{Parson} the easy axis results
from the competition between dipolar and quadrupolar interactions, which
depend on the temperature in different manner. In special situations a
surface nematic orientation temperature
dependent can be observed. The idea of
Parson has been generalized by
Sluckin and Poniewierski~\cite{Sluckin,S1,S2}
and by Sen and Sullivan~\cite{Sen}.
In all the models, the temperature surface transitions are due to a
temperature dependence of the anisotropic part of the anchoring energy,
which depends on the symmetry of the substrate and on the symmetry of
the nematic phase. 

The aim of our paper is to analyze the temperature dependence of the
anisotropic part of the surface energy. We assume that the nematic
liquid crystal is not polar. From the molecular point of view it has a
quadrupolar symmetry, whose principal axis coincides with ${\bf u}$. The
elements of the relevant tensor are
$q_{ij}=(3/2)[u_iu_j-(1/3)\delta_{ij}]$.     
In our analysis, we expand the surface energy
in series of spherical harmonic functions. The coefficients of the
expansion are the experimentally detectable anchoring coefficients.
According to our model, all the anchoring coefficients of the same order
depend on the temperature in the same manner. From this result it
follows that in nematic liquid crystals the alignment transitions driven by
the surface (the so called temperature surface transitions) are due to a
surface anchoring energy which contains contributions of different orders.

We analyze the temperature dependence of the anchoring
energy using an approach based on the mean field theory. In our analysis
we neglect all the inhomogeneities. We assume, furthermore that the
surface potential is short range. 

Let us consider a surface molecule of the nematic liquid crystal. It is
submitted to the mean field due to the other nematic molecules,
whose corresponding energy is $V_N$, and
to the interaction with the substrate, $V_S$.
In this framework, the total energy, $V$, of a given surface
molecule is $V=V_N+V_S$. If $V_N \sim V_S$ the
extrapolation length $b=K/W\sim a V_N/V_S$, where $K$ is an average
elastic constant and $a$ a molecular dimension, is of the order of a
molecular dimension~\cite{deGennes}.
In this case, in the continuum limit it is
possible to put $b=0$, and assume that the surface nematic orientation
is fixed by the surface interaction. This situation is known as the
strong anchoring case, and it is not interesting for us here. The interesting
case is the one in which $V_N\gg V_S$, corresponding to a situation
where $b\gg a$. This case corresponds to the weak anchoring situation,
to which we will limit our investigation. In our analysis the small
parameter used to expand the surface energy in power series is
$V_S/V_N\ll 1$, in the weak anchoring situation.
On the contrary, the surface scalar order parameter $S$ is
not supposed to be a small quantity.

$V_N$ describes the tendency of ${\bf u}$, which defines
the molecular orientation, to be oriented
along the nematic director ${\bf n}$.  Usually, it is 
approximated by means of the Maier-Saupe's mean field~\cite{Saupe},
$V_N^M$,
according to which
$V_N^M\propto n_i q_{ij} n_j=P_2({\bf n} \cdot {\bf u})$, where $P_2$ is
the second order Legendre Polynomial.
In this framework
$V_N^M=-v P_2({\bf n} \cdot {\bf u}) S$,
where $v$ is a molecular constant and
$S=\langle P_2({\bf n} \cdot {\bf u})\rangle$ the nematic scalar parameter.
A generalization of the Maier-Saupe theory has been proposed by
Humphries et al.~\cite{Humphries}. According to this generalized mean
field theory the nematic mean field is given by
\begin{equation}
\label{X}
V_N({\bf n} \cdot {\bf u})=
-\sum_l v_{2l}P_{2l}({\bf n} \cdot {\bf u})S_{2l},
\end{equation}
where $v_{2l}$ are molecular parameters, and
$S_{2l}=\langle P_{2l}({\bf n} \cdot {\bf u}) \rangle$ the
nematic order parameters, given by the self-consistent equations

\begin{equation}
\label{A-4}
S_{2l}=\frac {\int_0^1 P_{2l}({\bf n} \cdot {\bf u})
\exp [-\beta V_N({\bf n} \cdot {\bf u})] d({\bf n} \cdot {\bf u})}
{\int_0^1
\exp [-\beta V_N({\bf n} \cdot {\bf u})] d({\bf n} \cdot {\bf u})}.
\end{equation}
The Maier-Saupe potential,
$V_N^M$, is obtained from $V_N$ putting $v_{2k}=v \delta_{1,k}$.

The interaction connected to $V_S$
has to describe the tendency of the surface to orient
the surface nematic molecules along the ``easy direction", ${\bf n}_0$.
This direction
depends on the symmetry of the surface and on the molecular properties
of the mesophase. Since we limit our analysis to
non polar media, $V_S$ has to be an even function of ${\bf u}$. It follows
that $V_S$ is, actually, a function of the tensor ${\tensor q}$ and
can be written, in general, as
$V_S({\bf u})=V_S({\tensor q})=\sum_k w_k(0) L_k({\tensor q})$,
where $L_k({\tensor q})$ indicate the scalar quantities we can build with the
molecular tensor of elements
$q_{ij}=(3/2)[u_i u_j -(1/3)\delta_{ij}]$
and the elements of symmetry characterizing
the surface. Each term of the expansion of $V_S({\tensor q})$
represents a given
interaction, like induced dipole-induced dipole or
quadrupole-quadrupole and so on~\cite{Isra}; the ``intrinsic"anchoring
coefficients $w_k(0)$ are physical parameters connected
with the type of interaction described by $L_k({\tensor q})$.
$w_k(0)$ refer to specific
fundamental interactions, and are assumed to be
temperature independent. In this case thermal
effects arise only from the temperature dependence of the
degree of alignment of the nematic molecules. This conclusion is valid
only if in the temperature range of the nematic phase the physical
properties of the substrate can be considered constant. In the opposite
case $w_k(0)$ depend also on the temperature, via the substrate. Since
we assume that $w_k(0)$ are temperature independent, our
theory works well when the substrate is a solid crystal.
Deviations from our prediction are expected for nematic samples oriented
by means of surfactants. 

For our future considerations it is useful to describe the
molecular direction and the nematic director in terms of the
polar angles with respect to a cartesian
reference frame having the $z$-axis parallel to the geometrical normal
to the flat surface and the $x$-axis along the possible surface
anisotropy. Let $\Theta,\Phi$ and $\theta,\phi$ be  the polar and
azimuthal 
angles defining ${\bf u}$ and ${\bf n}$, respectively. Consequently  
\begin{equation}
\label{A0}
V_S({\tensor q})=V_S(\Theta,\Phi)= \sum_k w_k(0)L_k(\Theta,\Phi).
\end{equation}
By decomposing the functions $L_k(\Theta,\Phi)$ in series of spherical
harmonics functions $Y^m_k(\Theta,\Phi)$ we obtain
$L_k(\Theta,\Phi)=\sum_m a^m_k Y^m_k(\Theta,\Phi)$.
Since $L_k=L_k({\tensor q})$ and hence
$L_k(\Theta,\Phi)=L_k(\pi-\Theta,\pi+\Phi)$ for all $k$, we deduce that $k=2l$.
It follows that for non-polar nematic liquid crystals 
$L_{2l}(\Theta,\Phi) = \sum_m a^m_{2l} Y^m_{2l}(\Theta,\Phi)$, and the
microscopic surface energy can be written as
\begin{equation}
\label{A1}
V_S(\Theta,\Phi)= \sum_l w_{2l}(0)\sum_m a^m_{2l} Y^m_{2l}(\Theta,\Phi).
\end{equation}
The macroscopic anchoring energy
$W({\bf n})=W(\theta,\phi)$ is
obtained by averaging $V_S$ over the molecular orientations ${\bf u}$, or
over $\Theta$ and $\Phi$. Since in the problem under consideration
$V_S\ll V_N$, $V_S$ can be treated as a perturbation. According to the
thermodynamic perturbation theory~\cite{Landau}
we have $W(\theta,\phi)=\langle V_S(\Theta,\Phi) \rangle$, and hence,
as it follows from Eq.(\ref{A1}),
\begin{equation}
\label{A3}
W(\theta,\phi)=
\sum_l w_{2l} (0) \sum_ma^m_{2l} \langle Y^m_{2l}(\Theta,\Phi) \rangle, 
\end{equation}
where $ \langle A \rangle = Tr (\rho A) / Tr(\rho )$, and 
$ \rho = \exp ( -\beta V_N) $ is the density matrix. In order to derive
the macroscopic surface energy $W(\theta,\phi)$ we have first to express
$V_S(\Theta,\Phi)$ in terms of a polar coordinates system
based on the director
${\bf n}$ as polar axis. The cartesian reference frame has to be rotated
in such a way that  $ {\bf z}\,' ={\bf n}$. We will
indicate with $\vartheta,\varphi$ the polar angles of ${\bf u}$ with respect
to the rotated coordinate system. 
In this case~\cite{Rose}
\begin{equation}
\label{A4}
Y^m_l(\Theta,\Phi) = \sum_{m'}D^l_{m,m'}(\theta,\phi)
Y^{m'}_l(\vartheta,\varphi), 
\end{equation}
where $D^l_{m,m'}(\theta,\phi)$ are the elements of Wigner's matrix.
Since $\langle Y^{m'}_l(\vartheta,\varphi)\rangle=\langle Y^0_l(\vartheta)
\rangle \delta_{m',0}$ we obtain
from Eq.(\ref{A4})
$\langle Y^m_l(\Theta,\Phi)\rangle  =D^l_{m,0}(\theta,\phi)
\langle Y^0_l(\vartheta)\rangle$.
By taking into account that~\cite{Rose}
$D^l_{m,0}(\theta,\phi) = Y_l^m(\theta,\phi)$,
we have finally, as it follows from Eqs.(\ref{A0},\ref{A1},\ref{A3}), 
\begin{equation}
\label{A7}
W(\theta,\phi)
= \sum_l w_{2l}(0) S_{2l} L_{2l}(\theta,\phi),
\end{equation}
where we have taken into account that
$Y^0_{2l}(\vartheta,\varphi)=P_{2l}(\cos \vartheta)$.
Eq.(\ref{A7}) is a consequence of the fact that we regard all anisotropic
effects as perturbation, so that they do not need to be included in the
computation of the averages values. There is axial symmetry about the
direction of ${\bf n}$ in the imperturbed system and only the member
$m=0$ of the $Y_l^m$ is different from zero.
By comparing Eq.(\ref{A7}) with Eq.(\ref{A0}) we deduce that the
temperature dependence of the parameters describing the anisotropic part
of the surface energy is given by
\begin{equation}
\label{A8}
 w_{2l}(T) = w_{2l}(0) S_{2l}.
\end{equation}
This means that the temperature
dependence of $w_{2l}(T)/w_{2l}(0)$ coincides with the temperature dependence
of the $2l$-th scalar order parameter.

According to the analysis presented above, where the macroscopic
anchoring energy is given by the series expansion in spherical harmonic
functions shown in Eq.(\ref{A7}), the thermal renormalization
of the anchoring coefficients is given by Eq.(\ref{A8}).
>From these results it follows that the anchoring
coefficients of the same order in the expansion have the same
temperature dependence. Consequently, in the frame of our model,
temperature surface transitions 
are possible only in nematic samples whose anchoring energy contains
contributions from different order in the spherical harmonic functions
expansion. 

The ratios $S_{2l}/S \, vs.\, S$, for $l=2,3$
and $4$,  in the Maier-Saupe approximation, can be easily evaluated in the
nematic phase, where $0.4\leq S \leq 0.8$. A direct calculation shows that
$S_{2l}/S  \leq 0.2$, for $l=3,4$, as it is shown in Fig.1.
This explains why, usually, the anisotropic part of the surface
anchoring energy given by Eq.({\ref{A7}) is well approximated by few
terms~\cite{Paolo}.

In the low temperature region, where $-\beta V_N\gg 1$,
the fluctuations of ${\bf u}$ with respect to ${\bf n}$ are small. In
this region ${\bf n}\cdot{\bf u}=\cos \vartheta\sim
1-(1/2)\vartheta^2+{\cal O}(4)$, i.e. $\vartheta\ll 1$, and
$P_{2l}(\cos \vartheta)=1-[l(2l+1)/2]\vartheta^2 +{\cal O}(4)$.
Consequently, from Eq.(\ref{A-4}), the order parameter $S_{2l}$ is found
to be
\begin{equation}
\label{AB}
S_{2l}\sim 1-\frac{l(2l+1)}{B}\sim \exp\left\{-\frac{l(2l+1)}{B}\right\},
\end{equation}
where $B=\beta \sum_k k(2k+1) v_{2k} S_{2k}$. The main nematic scalar
order parameter $S=\langle P_2({\bf n}\cdot{\bf u})\rangle$ is given by
$S=\exp(3/B)$,as it follows from Eq.(\ref{AB}). The other order
parameters can be determined in terms of $S$ by $S_{2l}=S^{l(2l+1)/3}$.
In the low temperature region, the thermal renormalization of the
anchoring coefficient is then given by
\begin{equation}
\label{Ac}
w_{2l}(T)=w_{2l}(0)S^{\frac {l(2l+1)}{3}}.
\end{equation}
In particular, in this range of temperature, $w_2(T)/w_2(0)=S$ and
$w_4(T)/w_4(0)=S^{10/3}$.  The temperature dependence given by
Eq.(\ref{Ac}) reminds the Akulov-Zener law for magnetic anisotropy,
well known in ferromagnetism theory~\cite{Akulov,Zener}

As an example, we consider now a nematic liquid crystal limited by an
isotropic substrate.
In this case only the polar angle $\theta$ enters in the
description.
The analysis of the temperature surface transitions in a system of this
kind is usually performed by means of a Landau's expansion of the
anisotropic part of the surface energy~\cite{Sluckin,Gabbasova}.
According to this approach
$W({\bf n})$ is expanded in power series of the invariants made with the
elements of symmetry characterizing the nematic phase (which is the
nematic tensor order parameter of elements
$Q_{ij}=(3/2)S[n_in_j-(1/3)\delta_{ij}]$), and the substrate (which is
the geometrical normal ${\bf z}$). In the Landau-like approaches
the quantity playing the role of
expansion parameter is $S$. However, since the nematic-isotropic phase
transition is first order, $S$ is never very small (at the transition
point it is of the order of 0.3~\cite{PingSheng}). At the second order in
$S$, $W^L(\theta)=w_2^L P_2(\cos \theta)+w_4^L P_4(\cos \theta)
+{\cal O}(3)$, where $w_2^L=a_1S+a_2S^2$ and $w_4^L=a_3S^2$, in which
$a_1, a_2$ and $a_3$ are constant parameters, temperature
independent~\cite{Sluckin,Sen,Gabbasova}.

Now we want to compare the prediction of a Landau's expansion up to the
second order in $S$ with the result of our mean field analysis.  
In the case under consideration the angular functions $L_{2l}(\theta,\phi)$
reduce to $L_{2l}(\theta)=P_{2l}(\cos \theta)$, and
\begin{equation}
\label{e1}
W(\theta)=\sum_l w_{2l}(T) P_{2l}(\cos \theta).
\end{equation}
>From Eq.(\ref{A8}) we obtain
$w_2(T)/w_2(0)=S$.
This means that at the first order in $S$ the temperature dependence of
the anchoring energy deduced by means of symmetry considerations,
$w_2^L$, and by means of the mean field agree. 
However, for $l=2$ there is a discrepancy between the two
approaches. In fact, according to the mean field we have
$w_2(T)/w_2(0)=S$, and
$w_4(T)/w_4(0)=S_4 \neq S^2$,
whereas the Landau's  approach  predicts the temperature
dependencies $w_2^L(T)=a_1S+a_2S^2$ and $w_4^L(T)=a_3S^2$.
More precisely, it predicts a
renormalization of the coefficient of $P_2(\cos \theta)$, by means of a
$S^2$ contribution, and a temperature dependence of the coefficient of
$P_4(\cos \theta)$ like $S^2$. Of course, in the limit of small $S$ the
two predictions agree. In fact, if $S\ll 1$ the renormalization of
$P_2(\cos \theta)$ in $S^2$ can be neglected with respect to the linear
term in $S$. Furthermore, in this approximation,
$S_4 \propto S^2$. However, in the
case of large $S$ the discrepancy between the two approaches can be
large. In the low temperature region, where it is possible to use the
approximate expressions given by Eq.(\ref{Ac}) for the thermal
renormalization of the anchoring coefficients, our mean field
approach predicts $w_2(T)\propto S$ and $w_4(T)\propto S^{10/3}$. In
Fig.2 we show $S^2$, predicted by Landau-like models, and $S^{10/3}$,
predicted by our mean field theory in the low temperature region, ${\it
vs}$. $S$. As it is evident from this figure, our theory represents an
improvement with respect to the Landau-like approaches in the whole
temperature range.

To conclude we stress the main results reported in the paper.
We have shown that the
renormalization due to the thermal fluctuations of the anchoring coefficients
$w_{2l}$ is of the kind
$w_{2l}(T)/w_{2l}(0)=S_{2l}$ where
$S_{2l}$ is the $2l$-th scalar order parameter. In the particular case in
which the nematic phase is described by the Maier-Saupe theory,
$w_{2l}(T)/w_{2l}(0)$ coincides with the average value of the $2l$-th
Legendre polynomial. We have also shown that only at the lowest order in the
scalar order parameter the  simple approach based just on the symmetry
of the problem agrees with our mean field approach.  This is a
consequence of the hypothesis of small $S$, over which is based the
validity of the Landau-like expansions of $W({\bf n})$ in power of $S$.
We have proposed also approximate expressions for the thermal
renormalization of the anchoring coefficients, valid in the low
temperature region, where the fluctuations of the molecular directions
with respect to the nematic director are small.  

{\bf Acknowledgments} A.K.Z. has been partially supported by CNR-NATO
by means of a NATO Guest fellowships program. Many thanks are due to S.
Faetti, C. Oldano and S. Zumer for useful discussions. 

\begin{figure}
\label{1}
\caption{\protect$S_{2l}/S$ {\it vs}.$S$ according to Maier-Saupe theory. In
the usual nematic range, where $0.4\leq S \leq 0.8$,
\protect$S_6/S,\,S_8/S\leq 0.2$. This explains why, usually, two terms are
enough to approximate the macroscopic surface energy. The contribution
from higher harmonics disappears as a result of the thermal fluctuations
of the nematic molecules with respect to the director.}
\end{figure}

\begin{figure}
\label{2}
\caption{$S^2$ and $S^{10/3}$ {\it vs}.$S$. According to Landau-like
models at the second order in $S$, the thermal renormalization of the
surface energy in $P_4(\cos \theta)$ is proportional to $S^2$. According to
our mean field theory, it is proportional to $S_4$. In the low
temperature region, where the fluctuations of the molecular orientation
with respect to the nematic director are small, $S_4\sim S^{10/3}$. The
figure shows that our approximate expression for $S_4$ represents an
improvement with respect to the Landau-like analyses.}
\end{figure}

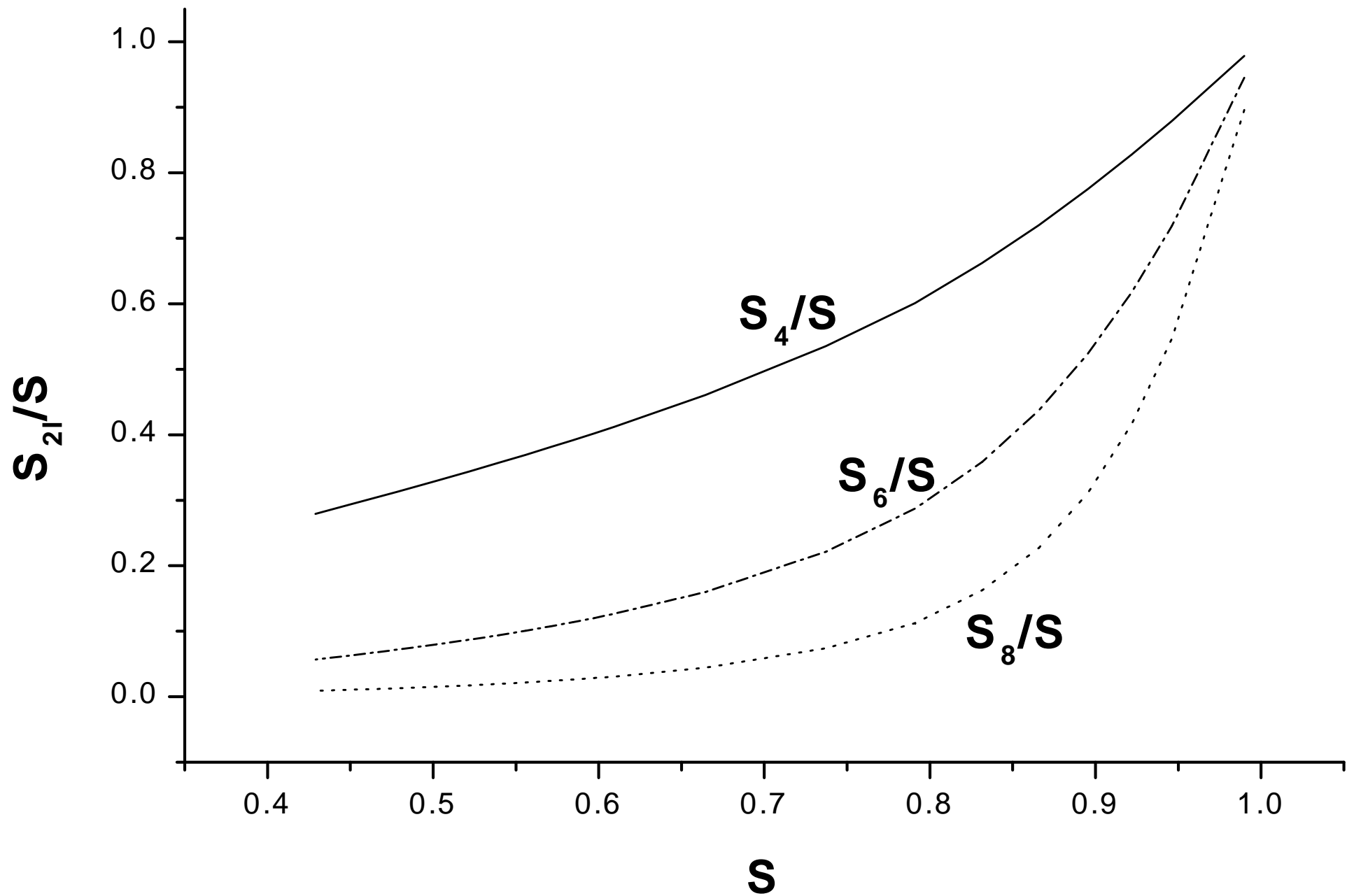
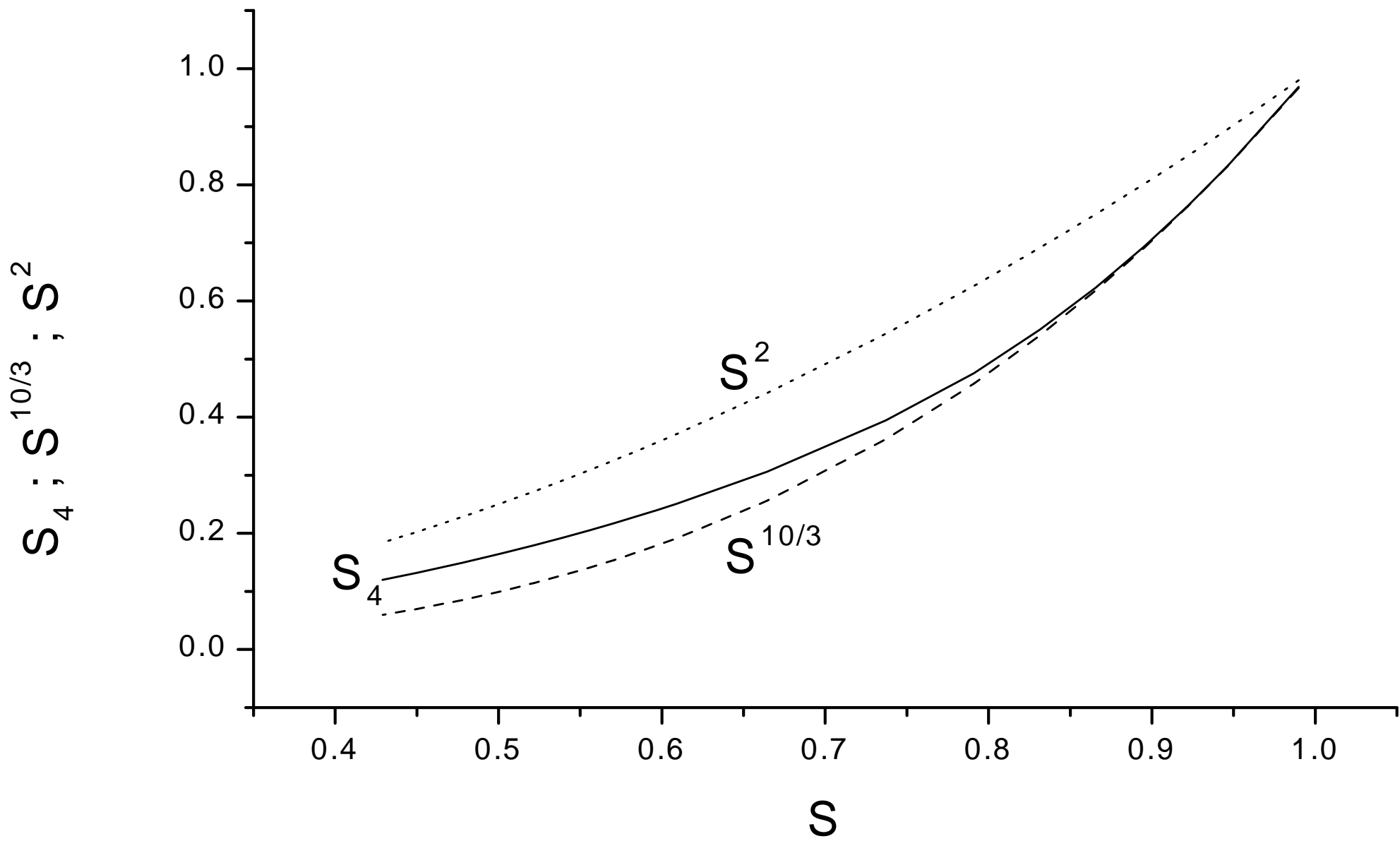
\end{document}